\documentclass[11pt,twoside]{article}
\usepackage{macro-fsut-eng}
\usepackage{graphicx}

\usepackage{ams math,amssymb,amsfonts,amsthm}
\usepackage[T1]{fontenc}

\newcommand{\be}{\begin{equation}}
\newcommand{\ee}{\end{equation}}
\newcommand{\ben}{\begin{displaymath}}

\newcommand{\een}{\end{displaymath}}

\newtheorem{Aux-Lemma}{Aux-Lemma}
\usepackage[T1]{fontenc} 

\usepackage{latexsym}
\usepackage{verbatim}
\begin{document}

\vskip 1.0cm
\markboth{Mar\'ia Eugenia Gabach Clement}{Shape of black holes}
\pagestyle{myheadings}

\vspace*{0.5cm}
\title{Shape of black holes}

\author{Mar\'ia Eugenia Gabach Clement}
\affil{FaMAF-UNC, IFEG CONICET, C\'ordoba, Argentina}

\begin{abstract}
It is well known that celestial bodies tend to be spherical due to gravity and that rotation produces deviations from this sphericity. 
We discuss what is known and expected about the shape of black holes' horizons from their formation to their final, stationary state.
 We present some recent results showing that black hole rotation indeed manifests in the widening
of their central regions, limits their global shapes and enforces their whole geometry to be close to the extreme Kerr horizon geometry at 
almost maximal rotation speed. The results depend only on the horizon area and angular momentum. 
In
particular they are entirely independent of the surrounding geometry of the spacetime and of the presence of matter satisfying the strong energy 
condition. We also discuss the the relation of this result with the Hoop conjecture.
\end{abstract}

\section{Introduction}
\label{sec:introduction}

This article is inspired by and based on a recent article [Gab2013] by MEGC and Martin Reiris. We refer the reader to 
[Gab2013] for further discussions and technical details.

The problem we wish to address is how to describe and characterize the shape of black holes. By this we mean the shape of their
horizons, how we can measure it, what  the restrictions on it are, if there are any, and how some physical parameters affect 
and determine this shape.


This problem has three important roots that we would like to refer to: The first one is the connection between the shape of a black hole and the shape of the matter configuration that collapsed
to form it. If general black holes were close to spherical, what would that say about highly non-
spherical collapse? One possibility would be that the deformations away from sphericity were in a sense, lost during the 
collapse that leads to the  black hole. Another possibility would be that non-spherical configurations would not collapse into a 
black hole at all, resulting in naked singularities or other compact configurations instead. In this sense, the allowed shapes of black holes 
might shed light into this very complicated collapse scenarios. The second root of the problem is the relation with the uniqueness theorems 
for the Kerr black hole and the basic questions here are the following. Are generic black hole solutions 
really that different from the Kerr black holes? Do they share any important property? Do they look similar in shape? The final point we want to 
remark is the connection between Newton's predictions on the shape of (non-relativistic) objects, with those of 
general relativity on black holes. We know that for small velocities and weak fields, general relativity reduces to Newton's 
gravitation. There are many relativistic phenomena that get lost in the Newtonian limit and we would like to understand if some of them manifest
though the shape of black holes but are absent in the shape of 'Newtonian' objects.

We will discuss these issues with a bit more detail through this article and a good place to start is to take a look at objects
that are more accessible to us than black holes, \textit{i.e.} stars and planets.


To a first approximation,  celestial objects are spherical. The main
reason for this is gravity, one expects that when enough mass is gathered close together, the resultant gravity will 
pull equally in all directions. Then, if there were no other effects present, the resulting shape 
would be a perfect sphere. But clearly this is not the case, as there are other ingredients involved, like the mass of the object, 
its rotation, the material it is made of, the magnetic fields, the surrounding fields and bodies. And these all combine
to produce  the different shapes we see in the sky. 

Of all these deformations away from sphericity, maybe the most common and easy to measure is a flattening due to rotation, resulting in 
configurations that become ever more oblate for increasingly rapid rotation. 
This effect, that we observe in the Sun, the Earth and most celestial bodies, can be described using Newtonian physics, with General Relativity playing no role. 

The natural questions are then: Can we expect the same behavior for black holes, the paradigm of relativistic objects? How do they look like?


As black holes can not be directly seen, the fine aspects of the shape of these objects can not be easily extracted from the images obtained by
telescopes. Therefore, we resort to theoretical models of realistic black holes.

\section{A model for realistic black holes}

One of the most used models to represent real black holes is the  solution of Einstein's equations found by Kerr [Ker1963]. It is stationary, 
vacuum, axisymmetric and asymptotically flat. It is characterized by two parameters, usually taken to be the  total ADM mass $m$ 
and total angular momentum $J$. For different values of these two parameters, we get completely different spacetimes. When $|J|>m^2$ we find a naked singularity. But when 
$J\leq m^2$ it gives a stationary, rotating black hole. Within the Kerr black hole family, when the angular momentum attains its maximum 
allowed value, $|J|=m^2$, the resulting black hole is called extreme Kerr black hole. The opposite case, that is, when the angular momentum 
is zero, leads to a static solution, the Schwarzschild black hole.

All through this article, we will refer to the Kerr black hole, with $|J|\leq m^2$, no naked singularities will be considered.

The location and properties of Kerr horizon can be easily read out from the explicit form of the metric (see, for instance, [Wal1984]). 
One can check that none of the rotating Kerr-horizons are exactly metrical spheroids.

Moreover, the Kerr black holes satisfy 
\begin{equation}\label{AJ}
8\pi |J|\leq A
\end{equation}
where $A$ is the horizon area. This inequality is relevant for the horizon description as it involves only quasi-local quantities that can be defined 
on the black hole horizon. Note that the equality in \eqref{AJ} corresponds to the extreme Kerr black hole, with $|J|=m^2=A/(8\pi)$. 
Therefore, for fixed area, the extreme Kerr black hole is the one that spins the fastest (within the family).

The problem of measuring the shape of Kerr's horizon and the deformations away from sphericity can be approached by computing the 
flattening coefficient due to rotation. It is defined as
\begin{equation}
 f:=1-\frac{C_{p}}{C_{e}}
\end{equation}
where $C_{p}$, and $C_{e}$ are respectively the lengths of the polar circle, that measures twice the distance between the poles, and the equatorial circle,
which is the greatest circle normal to the symmetry axis, see Figure 1.

\begin{figure}
\begin{center}
\hspace{0.25cm}
   \includegraphics[height=7cm, width=12cm]{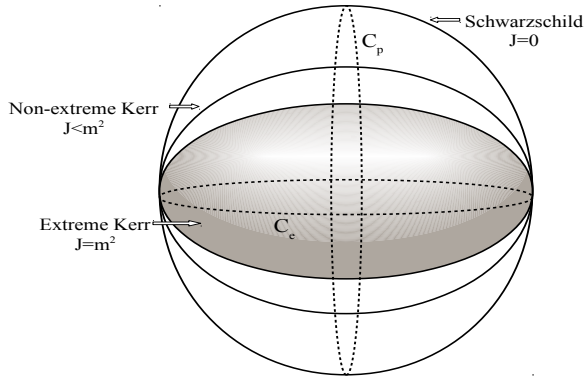}
\caption{Schematic representation of Kerr black hole horizons for equal values of mass and different values of the angular momentum. 
Also displayed are the quantities $C_p$ and $C_e$ for the Schwarzschild case. As the angular momentum increases from zero 
(Schwarzschild black hole) to the maximum value $|J|=m^2$ (extreme Kerr black hole) the horizon becomes 
ever more oblate.}
\end{center}
\end{figure}

For the Kerr family we find, by explicit computation, that $0\leq f\leq$0.36, and it  increases as the angular momentum 
increases (or the area decreases), which is in agreement with what is observed in stars and planets. The fastest the rotation, or the smaller
the object, the strongest the flattening. 
When there is no rotation at all (Schwarzschild black hole), the horizon is spherically symmetric, with zero flattening. Moreover, to the maximum 
value of $J$ corresponds the maximum flattening value, $f=0.36$, which is achieved by the extreme Kerr black hole. Keep in mind that a value of unity would mean a deformation to a disk.

To make contact with non-relativistic celestial bodies, note that the Sun has $f\sim10^{-5}$ and the Achernar star, the most flattened star known so far,
has $f= 0.17$ [Dom2003]. Quite remarkably, this last value coincides with the flattening of a Kerr black hole with $J\sim 0.88m^2$.

The above description shows that the Kerr family presents roughly a similar connection between rotation and flattening as most celestial bodies.
But what about more general black holes? Especially black holes that are not vacuum, stationary or axisymmetric like Kerr. Is the Kerr solution really that relevant in 
this more general scenario? In the next section we will see that Kerr black holes indeed turn out to be fundamental in the evolution of more complex
black holes and play a key role there.

\section{The final state}

Consider a spinning, very massive object that is collapsing to form a black hole. Independently of the very complicated and dynamical process that
takes place, the "black hole uniqueness theorems" say that once the dust has settled, and the system has calmed down, the final black hole is
simple [Bar1973-Chr1994]. It can be completely described in terms of a few 
parameters: the total mass, angular momentum and the electric charge of the black hole. Once these parameters are chosen, the system is 
determined completely. This striking property of black 
holes has been popularized as "the no-hair theorem" by Wheeler. The name alludes to the fact that 
only very few parameters are needed to describe those solutions, apart from the values of those parameters, black holes have no 
distinguishing characteristics (no "hair").

How do black holes become so simple?  The answer is gravitational waves. In general relativity, gravitational radiation escaping to infinity or
to the interior of a black hole carry almost all the complicated features present during the dynamical evolution stage. The gravitational field 
radiates away everything that can be radiated away, and the final stationary black hole is Kerr black hole, or its Kerr-Newman generalization when non-
vanishing electric charge is considered [Chr2012].

In summary, the uniqueness theorems give us detailed information about the final black hole and therefore, about the shape of its horizon, namely,
the final shape will be that of a member of Kerr family. But what about the beginnings? Right before the collapse?
And during the middle, dynamical stage? What do we know about black hole's shape then?

\section{The initial state}
One of the first results dealing with shapes and black holes is the Hoop conjecture, formulated  by 
Thorne in 1972 [Tho1972]. It reads ``{\it Horizons form when and only when a mass $m$ gets compacted onto a region whose 
circumference in every direction is less than or equal to $4\pi m$}". According to this conjecture, the circumference around the 
region must be bounded in every direction, and hence, a thin but long body of given mass would 
not necessarily  evolve to form a horizon. 

In principle, this conjecture talks about collapsing bodies, and not black holes. But if it were true, one would naively expect black holes not
to be very elongated, but localized in every direction, like the matter configuration they collapsed from. 

Unfortunately, the impreciseness in Thorne's conjecture had made this 
heuristic statement difficult to state, approach and prove. Since its 
formulation there has been a great amount of work making the idea more precise and attempting
to establish its correctness or otherwise [Sch1983-Sen2008-Gib2009]. 

The main problems with the original formulation of the Hoop conjecture have 
been nicely summarized by Senovilla [Sen2008] 
as follows:
In practice, it is not an easy task to determine the existence of the 
event horizon. 
An alternative has been the use of quasi-local, definitions of horizons, 
mainly apparent horizons and trapped surfaces.
Another problem is the notion of mass encircled by a hoop as there is no notion 
of a region encircled by a hoop. Moreover, one can pass arbitrarily small hoops 
around any distribution of 
matter by moving portions of the hoop relativistically.  
Finally, the uncertainty on the numerical constants used in the 
conjecture, several possibilities have appeared in the literature. 

Yet, despite all this, the Hoop Conjecture has been successful, in particular in spherical symmetry.  
Many numerical and analytical results have given it support (see [Sen2008] for details 
and full references). The problems in the original formulation and
the subsequent studies have not only given insight as to what to expect at the initial stage in the black hole evolution, but also inspired 
and guided the technical aspects in the study of shape of black holes.

In  2013, Gabach Clement and Reiris study the shape of dynamical, axisymmetric, rotating black holes and use measures solely
in terms of the area and angular momentum of the black hole horizon.

\section{Dynamical stage}

We discuss here some of what is known and the expectation about the properties of the shape of black hole horizons in the middle 
stage, namely, after the black hole formation and before the stationary, final phase.

There are two fundamental issues one must resolve in order to describe the shape of a black hole. First is how one will represent the black hole
horizon, and second, how its shape will be measured. As to the first issue, apparent horizons or marginally outer trapped surfaces have been the
preferred choice during the last years [And2008, And2009]. As to the measures of shape, one possibility is to find a background, well known configuration 
to compare with. Another possibility would be to construct coefficients, like the flattening mentioned in connection with the Kerr back hole 
that give an intrinsic notion of deformation. Finding a well defined and practical notion is an important and complicated
point that, as we will show in Section \ref{Axial}, has a straightforward solution in axial symmetry.

Gibbons [Gib2009-Gib2012] considers apparent horizons as representations of black hole horizons and studies two measures of shape: the length 
of the shortest non trivial closed geodesic $\ell$ and the Birkhoff's invariant $\beta$. Remarkably he finds that if the surface admits an 
antipodal isometry and that Penrose inequality holds, then $\ell\leq 4\pi m$. Gibbons goes further and 
conjectures that $\ell\leq\beta\leq 4\pi m$ hold in the general case, without antipodal symmetry.

\begin{figure}
\begin{center}\label{Horizon}
\hspace{0.25cm}
   \includegraphics[height=7cm, width=12cm]{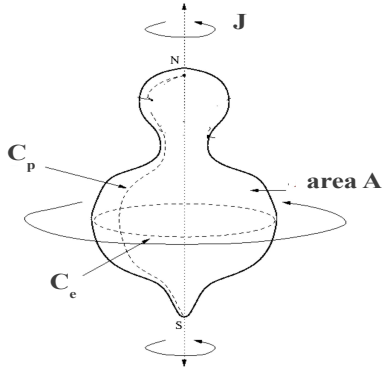}
\caption{Schematic representation of an axisymmetric black hole horizons, together with the great circle of length $C_e$ and the 
polar circle of length $C_p$ (image partly taken from [Gab2013]).}
\end{center}
\end{figure}

\subsection{Axial symmetry}\label{Axial}

Gabach Clement and Reiris [Gab2013] study axially symmetric black holes and represent their horizons by stable marginally outermost trapped 
surfaces. These surfaces are such that the outgoing null expansion is zero. The stability property is crucial and plays a central role in many 
features of black-holes, in particular, the horizon's shape. 
In axial symmetry the study of shape of the horizon is somewhat simplified because the symmetry axis defines two 
meaningful hoops on the surface. One is given by the greatest meridian and the other, by the great circle or greatest 
axisymmetric orbit, see Figure 2. In analogy with the analysis for Kerr black hole, denote the lengths of these curves by $C_{e}$ and $C_{p}$ 
respectively

In rotating, axially symmetric black holes, the extreme Kerr black hole plays a key role. This can be in part seen from the following inequality
\be\label{AJgeneral}
8\pi |J|\leq A
\ee
valid for \textit{all} dynamical, axially symmetric black holes [Hen2008, Ace2011, Dai2011, Jar2011, Gab2013a]. Note that \eqref{AJgeneral} looks exactly the same as the inequality \eqref{AJ} 
we described for the Kerr family. Moreover, the extreme Kerr black hole saturates \eqref{AJgeneral} among axially symmetric black holes. Therefore
one can think of this black hole as the smallest one for a given angular momentum, or the most rapidly rotating one for given size. These observations
suggest that extreme Kerr black hole may be a good candidate to compare the shape of a general black hole with.

The study of the shape of black holes in [Gab2013] is done through the comparison with extreme Kerr black hole
and the estimation of the horizon's lengths $C_p, C_e$ and the flattening coefficient $f$. The main results of that article are discussed below.

\vspace{0.5cm}

\textit{Bounds on $C_e, C_p$ and $f$.}

The most noticeable effect of rotation is a thickening of the bulk of the horizons. More precisely, the 
length $C_e$ of the great circle is subject to the lower bounds 
 \begin{equation}\label{THICKBH1}
  \frac{16\pi|J|^2}{A}\leq\frac{8|J|}{\delta+\sqrt{\delta^2+4}}\leq\left(\frac{C_{e}}{2\pi}\right)^2,
 \end{equation}
 where
\be\label{delta}
\delta=2\sqrt{\bigg(\frac{A}{8\pi|J|}\bigg)^{2}-1}.
\ee
These formulae say that rotation imposes a minimum (non-zero) value for the length of the greatest circle. For a given angular momentum, the horizon 
can not be too thin, but it gets thickened perpendicular to the rotation axis. Nevertheless they do not 
say whether the greatest circle lies in the 'middle region' of the horizon or 'near the poles', nor does it say anything about the size 
of other axisymmetric circles. 

\vspace{0.5cm}

On the other hand, there is an upper bound on the length of the great circle, given by
\be \label{THICKBH3}
 \left(\frac{C_{e}}{2\pi}\right)^2\leq4|J|\frac{\delta+\sqrt{\delta^2+4}}{2}\leq\frac{A}{\pi}
 \ee
showing that ultimately, the area controls the maximum size of the great circle. Note that for a perfect sphere, 
$\left(\frac{C_{e}}{2\pi}\right)^2=A/(4\pi)$, and more importantly, for extreme Kerr black hole, $\left(\frac{C_{e}}{2\pi}\right)^2=A/(2\pi)$, giving 
a factor of 2 with the right hand side of \eqref{THICKBH3}. 

Putting \eqref{THICKBH1} and \eqref{THICKBH3} together, one sees that they coincide when $\delta=0$ at $C_e/2\pi=2\sqrt{|J|}$ which is the value for 
the extreme Kerr horizon. This is not a coincidence and will be discussed below.

The final relation that is presented, is connected with the flattening factor $f$ and reads
 \begin{equation}\label{THICKBH2}
f:=1-\frac{C_p}{C_e}\leq 1-\frac{1}{\sqrt{2}\pi}\sim0.77
 \end{equation}
which shows that stable rotating horizons of a given area $A$ and angular momentum $J\neq 0$, cannot be arbitrarily 
oblate or 'thick'. Let us pause a moment to analyze this bound. Recall that the extreme Kerr black hole has $f_{eKerr}=0.36$ which is roughly 
half right hand side of \eqref{THICKBH2}.
As extreme Kerr black hole is the black hole that, given the horizon area, rotates the fastest (because it is the only axisymmetric solution that
saturates the bound \eqref{AJgeneral}), then one would expect that it is the most flattened black hole. Therefore, naively we expect $f\leq f_{eKerr}$ for all
axisymmetric black holes. Nevertheless, we get twice that value in \eqref{THICKBH2}, which leaves room to improve.

\vspace{0.5cm}

{\it Rotational stabilization}. Rotation stabilizes the shape of stable horizons in such a way that for given A and J their entire shapes are controlled 
 (not just $C_{p}$ or $C_{e}$). 
 \begin{equation}
 \|g-g_{eKerr}\|_{C^{0}}\leq F(\delta)
\end{equation}
for a certain finite function $F(\delta)$, where $g$ is the 2-metric on the horizon, $g_{eKerr}$ is the 2-metric on the extreme Kerr horizon with 
angular momentum $J$ 
and $\delta$ was defined in \eqref{delta}. In [Gab2013] it is shown that not only the metric (and therefore, the whole geometry of the horizon) is 
controlled in that way, but also the rotational potential are completely controlled by $A$ and $J\neq 0$. 
It also shows that stable holes with $A/8\pi|J|$ close to one must be close to the extreme Kerr horizon.

\vspace{0.5cm}

\textit{Enforced shaping} 

At very high rotations all the geometry of the horizon tends to that of the extreme Kerr horizon, regardless of 
the presence of any type of matter, as long as it satisfies the Strong Energy Condition:

\vspace{0.5cm}

\textit{Hoop like inequality}.

The following result makes contact with Thorne's Hoop conjecture. Consider a stable, axisymmetric, outermost minimal surface\footnote{See [Gab2013a] for a 
connection between stable axisymmetric minimal surfaces and stable axisymmetric trapped surfaces in the context of geometrical inequalities involving angular momentum and area.} on a 
maximal axisymmetric and asymptotically flat initial data for Einstein equations. Matter satisfying the strong energy condition is also allowed.
Then, assuming that Penrose inequality $16\pi m^2\geq A$ [Mar2009] holds, it is deduced that the length of the great circle of the minimal surface 
satisfies
\begin{equation}\label{newHoop}
 C_{e}\leq 8\pi m,
\end{equation}
where $m$ is the ADM mass.

Note that again, there is  a factor of 2 (as in \eqref{THICKBH2}) on the right hand side with respect to the Thorne conjectured value of $4\pi m$.

Finally, regarding Gibbons' conjecture on the Birkhoff's invariant mentioned earlier, $\beta\leq 4\pi m$, using the above results, in
[Gab2013]  it is proved that for axisymmetric outermost minimal spheres one has $\beta\leq C_e$ and therefore from \eqref{newHoop}, 
$\beta\leq 8\pi m$. Whether $8\pi$ instead of Gibbon's $4\pi$ is the right coefficient for $m$ is not known. Nevertheless, the $8\pi$ factor in 
\eqref{newHoop} is not sharp. If one expects the 
Penrose inequality to hold also for apparent horizons, then the argument before would 
work the same and one would obtain $C_e\leq 8\pi m$ as well.

\section{Final comments}

There are many open problems that still need attention.  

Regarding the Hoop conjecture, there have been works on establishing a precise formulation and extensions to other theories, higher dimensions  and 
special spacetimes [Sen2008-Yos2008-Gib2009-Khu2009-Yoo2010-Muj2012], nevertheless, a clear and general statement and proof is lacking. 

With respect to the shape of black holes, one would like to take into account other ingredients like matter content and type and the influence of
magnetic fields. Another issue of great relevance is the condition of axial symmetry imposed in [Gab2013]. In the absence  of this symmetry, as we mentioned before, the measures of shape 
and the study of the horizon geometry become much more complicated, as is manifest in the work of Gibbons [Gib2012]. But even within axial symmetry, 
as it was mentioned, there is still room for improvement in the bounds found in [Gab2013].

Finally, it would be interesting to analyze what general relativity has to say about the shape of material objects, instead of black holes, in the 
spirit of [Sch1983].

\acknowledgments It is a pleasure to thank the organizers of the \textit{Gravitation, Relativistic Astrophysics and Cosmology
Second Argentinian-Brazilian Meeting} that took place in Buenos Aires in April 2014 for their kind invitation and hospitality.


\begin{references}

\reference  [Ace2011] Andres Acena, Sergio Dain, and Maria E. Gabach Clement. Horizon area-angular 
momentum inequality for a 
class of axially symmetric black holes. Class.Quant.Grav., 28:105014, 2011.

\reference [And2008] Lars Andersson, Marc Mars, and Walter Simon. Stability of marginally outer trapped
surfaces and existence of marginally outer trapped tubes. Adv. Theor. Math. Phys., 12, 2008.

\reference [And2009] Lars Andersson and Jan Metzger. The Area of horizons and the trapped region. Com-
mun.Math.Phys., 290:941-972, 2009.


\reference [Bar1973] J. M. Bardeen, Brandon Carter and Stephen W. Hawking. The fou laws of black hole mechanics. Commun. Math. Phys. 31 (1973), no. 2, 161-170. 

\reference [Bra2001] Hubert L. Bray. Proof of the Riemannian Penrose inequality using the positive mass
theorem. J. Differential Geom., 59(2):177–267, 2001.

\reference [Chr1994] Piotr Chrusciel. 'No Hair' Theorems - Folklore, Conjectures, Results. Contemp.Math.170:23-49,1994 


\reference [Chr2012] Piotr T. Chrusciel, Joao Lopes Costa and Markus Heusler. Stationary Black Holes: Uniqueness and Beyond. Living Rev. 
Relativity 15 (2012), 7.

\reference [Dai2011] Sergio Dain and Martin Reiris. Area-Angular-Momentum Inequality for Axisymmetric Black
Holes. Physical Review Letters, 107(5):051101, 2011.

\reference [Dom2003] Armando Domiciano de Souza, P Kervella, S. Jankov, L. Abe, F. Vakili, et al. The
Spinning-top Be star Achernar from VLTI-VINCI. Astron.Astrophys., 407:L47-L50,
2003.

\reference [Gab2013] Maria E. Gabach Clment and Martin Reiris. Shape of rotating black holes
Phys. Rev. D 88, 044031, 2013

\reference [Gab2013a] Maria E. Gabach Clement, Jose Luis Jaramillo, and Martin Reiris. Proof of
the area-angular momentum-charge inequality for axisymmetric black holes.
Class.Quant.Grav., 30:065017, 2013.

\reference [Gib2009] Gary W. Gibbons. Birkhoff's invariant and Thorne's Hoop Conjecture. 2009.



\reference [Gib2012] Gary W. Gibbons. What is the Shape of a Black Hole? AIP Conf.Proc., 1460:90-100, 2012.

 


\reference [Hen2008] Jorg Hennig, Marcus Ansorg, and Carla Cederbaum. A Universal inequality between
angular momentum and horizon area for axisymmetric and stationary black holes
with surrounding matter. Class.Quant.Grav., 25:162002, 2008.


\reference [Jar2011] Jose L. Jaramillo, M. Reiris, and S. Dain. Black hole area-angular-momentum inequality in nonvacuum spacetimes. Physical Review Letters D, 84(12): 121503, December
2011.

\reference [Ker1963] Roger P. Kerr. Gravitational field of a spinning mass as an example of algebaically special metrics. Phys. Rev. Lett. 11,
237-238, 1963.

\reference [Khu2009] Marcus A. Khuri. The Hoop Conjecture in Spherically Symmetric Spacetimes. Phys.Rev.D80:124025,2009.

\reference [Mar2009] Marc Mars. Present status of the Penrose inequality. Class.Quant.Grav.26:193001, 2009. 

\reference [Muj2012] A. H. Mujtaba, C. N. Pope.  The Hoop Conjecture for Black Rings.  arXiv:1211.6030.
   
    
\reference [Sen2008] Jose M M. Senovilla. A Reformulation of the Hoop Conjecture. Europhys.Lett.81:20004, 2008. 

\reference [Sch1983] Richard Schoen and S.T. Yau. The Existence of a Black Hole Due to Condensation of Matter. Commun. Math. Phys. 
90, 575-579 (1983).

\reference [Tho1972] Kip S. Thorne. Nonspherical gravitational collapse: A short review. In J.R. Klauder, ed-
itor, Magic Without Magic: John Archibald Wheeler. A Collection of Essays in Honor
of his Sixtieth Birthday, pages 231-258. W.H. Freeman, San Francisco, 1972.

\reference [Yoo2010] Chul-moon Yoo, Hideki Ishihara, Masashi Kimura, Sugure Tanzawa. Hoop Conjecture and the Horizon Formation Cross-Section 
in Kaluza-Klein Spacetimes. Phys.Rev.D81:024020,2010 

\reference [Yos2008] Hirotaka Yoshino. Highly distorted apparent horizons and the hoop conjecture. Phys.Rev.D77:041501,2008.

\reference [Wal1984] Robert Wald. General Relativity. The University of Chicago Press.  ISBN 0-226-87033-2. 1984.

\end{references}
\end{document}